\documentclass[12pt]{iopart}

\usepackage{iopams}  
\usepackage{graphicx}

\begin{document}

\date{\today}

\title{Fast ignition of fusion targets by laser-driven electrons}

\author{J.J.~Honrubia}
\address{
E.T.S.I. Aeron$\acute{a}$uticos, Universidad Polit$\acute{e}$cnica de Madrid, Spain}
\ead{javier.honrubia@upm.es}

\author{J.~Meyer-ter-Vehn}
\address{
Max-Planck-Institut f\"ur Quantenoptik, Garching, Germany}
\date{\today}

\begin{abstract}
We present hybrid PIC simulations of fast electron transport and energy
deposition in pre-compressed fusion targets, taking full account of
collective magnetic effects and the hydrodynamic response of the
background plasma. Results on actual ignition of an imploded fast
ignition configuration are shown accounting for the increased beam
divergence found in recent experiments [J.S. Green et al.,
Phys. Rev. Lett. {\bf 100}, 015003 (2008)] and the reduction
of the electron kinetic energy due to profile steepening
predicted by advanced PIC simulations [B. Chrisman et al.
Phys. Plasmas {\bf 15}, 056309 (2008)].
Target ignition is studied as a function of injected electron energy,
distance of cone-tip to dense core, initial divergence and kinetic energy
of the relativistic electron beam. We found that beam collimation
reduces substantially the ignition energies of the cone-guided
fuel configuration assumed here.

\end{abstract}


\section{Introduction}

Fast ignition \cite{tabak1,tabak2} involves transport of GA currents
of laser-driven electrons through dense coronal plasma of imploded
fusion targets. Recent studies \cite{honrubia4, honrubia5} have shown that
the beam can undergo the resistive filamentation instability
when passing through coronal plasma and that beam collimation
by self-generated fields can reduce ignition energies
substantially. These results were obtained for rather focused
beams (initial divergence half-angle = 22$^{\circ}$)
with electron kinetic energies given by the ponderomotive scaling.
Recent experiments \cite{Green, Lancaster} have evidenced an increase
of beam divergence with laser intensity, e.g. electron
effective propagation angles of 35$^{\circ}$ have
been reported for fast ignition
relevant conditions. In addition, advanced
PIC simulations \cite{sentoku2008} have shown that the
ponderomotive scaling overestimates fast electron kinetic
energies due to the plasma electron density profile steepening
caused by photon pressure at laser irradiances of the
order of 10$^{20}$ W/cm$^2$.
In this paper, we broaden the scope of our previous studies
\cite{honrubia4, honrubia5} and present results on actual ignition
of an imploded fast ignition configuration accounting for the
increase of the beam divergence and the reduction of the electron
kinetic energy mentioned above. In addition, we assume a supergaussian
distribution in radius of the beam electrons to enhance azimuthal
magnetic field generation and beam collimation. An imploded cone-guided
fuel configuration is considered, and target ignition is studied as
a function of injected electron energy,
distance of cone-tip to dense core, initial divergence and
mean kinetic energy of the relativistic electron beam. The present
study has been motivated by the future fast ignition facilities
such as HiPER \cite{dunne}.

In the hybrid approach pursued here, the relativistic electron beam is
treated by 2D/3D PIC including collisional energy loss, while the
high density background plasma is modelled by resistive MHD equations
including hydrodynamic motion to describe magnetic field suppression by
plasma return currents. Full scale kinetic simulation may become
possible in the future \cite{sentoku2008}, but presently the hybrid
approach pursued here offers a unique option to investigate important
transport features such as current filamentation and magnetic beam
collimation simultaneously with ignition physics (fusion reactions,
$\alpha$-particle transport and deposition, thermal radiation
transport, hydrodynamics and heat conduction). One may recall
that, so far, most fast ignition simulations \cite{solodov2007}
assumed ballistic straight-line beam transport, neglecting all
the intricacies of high-current (GA) transport in plasma.

The present paper is organised as follows. In Section 2, the simulation
model and the initial distribution function assumed for the
fast electron beam are described. In Section 3, the imploded fuel
target configuration and the beam parameters used in our study
are presented. In Section 4, we analyse the fast electron energy
deposition in the target. In Sections 5 and 6, the coupling
efficiencies and the ignition energies are parametrically
studied, respectively, as a function of cone-tip to dense
core distance, initial divergence and mean kinetic energy
of the fast electron beam. Finally, in Section 6 we summarise
our results.

\section{Simulation model}

We used the hybrid model proposed by Bell \cite{bell} and further
developed by Davies \cite{davies} and Gremillet et al.
\cite{gremillet}. This model is adequate for describing self-magnetized
transport in high-density fuel, where kinetic energy transfers
and most of the beam-plasma instabilities are suppressed by
collisions. It treats only the relativistic
beam electrons by PIC and models the background plasma by the
return current density {\bf j$_r$}, tied to the electric field
{\bf E} by Ohm's law with resistivity $\eta$. Maxwell's equations
are used in the form
\begin{eqnarray}\label{e5}
   {\bf E} & = & \eta {\bf j_{r}},  \\
   {\bf j_{b}}+{\bf j_{r}} & = & \frac{1}{\mu_0} \nabla \times {\bf B}, \\
   \partial {\bf B} / \partial t &=& - \nabla \times {\bf E},
\end{eqnarray}
where {\bf j}={\bf j$_b$}+{\bf j$_r$} is the net current density, defined
as the sum of beam and return current densities. The displacement
current and charge separation
effects can be neglected since in this high-density environment
relaxation times and Debye lengths are much smaller than the
sub-picosecond and micrometer scales in which we are interested here.
The recent study by Kemp et al. \cite{kemp2} justifies to
choose resolution less than the Debye and skin lengths at
high densities due to the suppression of collective
effects by collisions. In the present model, the beam deposits
energy into plasma electrons by direct classical Coulomb deposition
and via return current Ohmic heating with power density $\eta j_r^2$.
Our hybrid-PIC approach reproduces cone-target experiments \cite{kodama1, kodama2},
similar to previous hybrid simulations \cite{campbell, mason}.
Since laser pulses of about 15 ps are needed to ignite the pre-compressed
fuel, we have included in our model thermal electron conduction, multigroup
radiation transport, fusion reactions, alpha particle transport and energy
deposition, and hydrodynamics. This integrated model
is used here in 2D r - z cylindrical geometry mode. It is worth pointing
out that previous hybrid simulations \cite{honrubia4,campbell,mason}
assumed a stationary background plasma while in the present model
plasma motion is described by radiation-hydrodynamic equations. The
details of the fast electron transport code can be found in
\cite{honrubia2,honrubia1,honrubia3}.

We assume in the present calculations that
beam filamentation at the cone tip
\cite{mason, Sentoku2004} or Weibel and
two-stream instabilities \cite{silva,bret} in
the low density zones do not perturb
beam propagation from the critical
surface to the minimum density of 2 g/cm$^3$
considered in this paper. This is supported by the cone
target experiments carried out by Kodama et al. \cite{kodama1, kodama2}
and also by the recent PIC calculations by Chrisman, Sentoku and Kemp
\cite{sentoku2008} showing that, even though the fast
electron beam is split into filaments at the cone tip,
these filaments are strongly damped in the dense plasma
beyond the cone.

Because the ultra-intense laser pulse interaction with
the cone cannot be described by hybrid codes, we rather
model the injected beam in form of a directed energy distribution
typically fitted to PIC simulations \cite{sentoku2008,pukhov}
or experiments \cite{kodama1,kodama2,nakamura2008}. The energy
distribution of beam electrons is assumed to be one-dimensional (1D)
relativistic Maxwellian with temperature
$kT_b = f_{T} \times m_ec^2\,[(1+I\lambda^2/1.37\times 10^{18})^{1/2}-1]$
obtained for local laser intensity $I$ (in units of W/cm$^2$) and
wavelength $\lambda$ (in $\mu$m). Here, $f_{T}$ is a parameter
that accounts for the plasma electron density profile steepening
due to the laser ponderomotive force reported in Ref. \cite{sentoku2008}.
We assume $f_T$ = 0.80, 0.67 and 0.51 to obtain electron mean
kinetic energies of 2.0, 1.6 and 1.2 MeV for a laser intensity of
2$\times$10$^{20}$ W/cm$^2$ and $\lambda$ = 0.53 $\mu$m,
respectively\footnote{For electron temperatures kT$_b$ =
2.24, 1.88 and 1.43 MeV, the 1D relativistic Maxwellian
distribution gives $\langle E \rangle$/kT$_b$ $\approx$
0.88, 0.86 and 0.84, respectively.}.
We have chosen the 1D relativistic Maxwellian distribution due to the
1D character of the electron acceleration by ultra-intense laser pulses.
In addition, the lack of experimental data about spectra of electron beams
produced by laser pulses of a few times $10^{20}$ W/cm$^2$ and 10 - 20 ps
duration also supports our choice. It is worthwhile pointing out that
the mean energy of the 1D relativistic Maxwellian distribution
($\langle E \rangle_{1D} = kT$ in the ultra-relativistic limit)
is substantially lower than that corresponding to the 3D relativistic
Maxwelian one ($\langle E \rangle_{3D} = 3kT$ in the ultra-relativistic
limit) for the same laser intensity, leading to beam electrons with
lower kinetic energy and penetration depth. 

Initial divergence angle of the relativistic
electron beam is a crucial parameter for fast ignition.
Recent experiments of hot electron generation and transport
in solid foil targets with
$I \lambda^2$ ($\sim 4 \times 10^{19}$ W$\mu$m$^2$/cm$^2$ )
and pulse durations $\tau$ ($\sim$ 5 ps)
relevant for fast ignition have evidenced beam divergence
angles of about 35$^{\circ}$ \cite{Green, Lancaster}.
Our hybrid simulations can reproduce this 
result by assuming an initial divergence
half-angle $\langle \theta \rangle$ around 50$^{\circ}$. The
difference between this initial half-angle 
and the effective propagation full-angle of 35$^{\circ}$
is due to the beam collimation
by self-generated azimuthal magnetic fields
$B_{\theta}$, which strength increase with
pulse duration as shown by equation (3).
Integrated simulations for initial
half-angles $\langle \theta \rangle \ge$ 50$^{\circ}$ give
electron-beam ignition energies $E_{ig} >$ 50 kJ,
well over the energies envisaged for the
future facilities \cite{dunne}. Fast electron
generation in cones may reduce the beam divergence
due to surface acceleration and focusing toward
the cone tip \cite{nakamura2004,nakamura2007}. However,
there are experimental evidences showing almost
no effect of the cone on beam focusing \cite{Baton}.
Here, we have considered the initial beam opening
half-angle as a parameter, varying from the 22$^{\circ}$ measured
by Kodama et al. \cite{kodama1,kodama2} to the 50$^{\circ}$ used to
reproduce the experiments mentioned above.
The initial beam divergence is modelled as follows:
a fast electron with kinetic energy $E=(\gamma-1)m_ec^2$
is randomly injected in a cone with a half-angle
given by the ponderomotive scaling formula
$\theta = \tan^{-1}\{f_{\theta}\times[2/(\gamma-1)]^{1/2}\}$,
where $\gamma$ is the Lorentz factor of the electron and
the parameter $f_{\theta}$ is used to adjust the initial
divergence half-angle of the whole electron beam.
We have chosen $f_{\theta}$ = 1 for the reference case that
corresponds to $\langle \theta \rangle =35^\circ$, well below the
initial divergence required to reproduce the experiments of
Refs. \cite{Green, Lancaster}, but substantially higher than
the $\langle \theta \rangle =22^\circ$ considered in previous
studies \cite{honrubia4}. 

\begin{figure}
!\hspace{2.5cm}
\begin{center}
\includegraphics[width=0.80\textwidth]{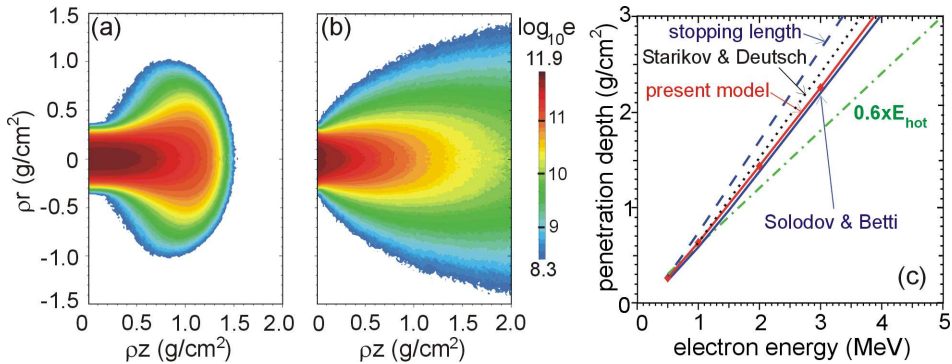}
\end{center}
\caption{\label{fig:1} Energy deposition
isocontours of (a) monoenergetic 2 MeV electrons
impinging perpendicularly on a DT slab of
400 g/cm$^3$, and (b) electrons with the spectral
distribution explained in the text, mean energy
$\langle E \rangle$ = 2 MeV and divergence
half-angle $\langle \theta \rangle= 22^\circ$ (HWHM).
Colour isocontours are in logarithmic scale in units
of J/cm$^3$. (c) Comparison of the electron penetration
depth obtained by different models.}
\end{figure}

Fast electron collisions with the background
DT plasma are modelled by the relativistic
Fokker-Planck
equation with standard Coulomb cross sections.
The Fokker-Planck equation is solved by the
Monte Carlo method \cite{Salvat}.  The energy
deposition of a monoenergetic 2 MeV electron
beam with a supergaussian profile in radius
and a half width at half maximum (HWHM) of
20 $\mu$m impinging perpendicularly on a
400 g/cm$^3$ DT slab is shown in figure~\ref{fig:1}.
The beam is collimated at half of its range,
being subject to scattering and blooming in
the second half, as reported in \cite{Li}.
The fast electron penetration depth
obtained with the present model is similar to
that recently published by Solodov and Betti
\cite{solodov} and also to the quantum calculations
by Starikov and Deutsch \cite{deutsch, deutsch1}
and about 20\% larger than the scaling
$\rho R$ [g/cm$^2$]= 0.6 $\times E_{hot}$ [MeV]
\cite{atzeni3}, as shown in figure~\ref{fig:1}(c).
It is worth noticing that the energy deposition
changes significantly when beams with realistic
energy and angular distributions are considered.
For instance, in figure~\ref{fig:1}(b) the energy
deposition is shown for the same beam as in
figure~\ref{fig:1}(a), but with a mean divergence
half-angle of 22$^{\circ}$ (within the HWHM of
the super-Gaussian profile in radius) and a
1D relativistic Maxwellian distribution with
mean energy $\langle E \rangle$ = 2 MeV. Notice the
apparent range lengthening due to the beam
energy spectrum and the subsequent delocalization of
electron energy deposition.

Regarding the energy transport models, we have used
one-group flux-limited thermal electron conduction and
multigroup radiation transport. Classical Spitzer
electrical and thermal conductivities are chosen
for the DT plasma with the Coulomb logarithms given
in Ref.~\cite{NRL}. Degeneracy effects have not been taken
into account in the transport coefficients because
the DT is heated to keV temperatures in a few ps and
field generation is not important in the dense core,
as will be discussed in Section 4. We have neglected
also the effects of self-generated magnetic fields on
transport coefficients. MPQeos tables \cite{kemp} have
been used to compute electron and ion temperatures from
the deposited energy. Alpha-particle transport and energy
deposition calculations have been performed by a
3D Monte-Carlo model with standard stopping power and
mean deflection coefficients. 

\begin{figure}
!\hspace{4cm}
\begin{center}
\includegraphics[width=.65\textwidth]{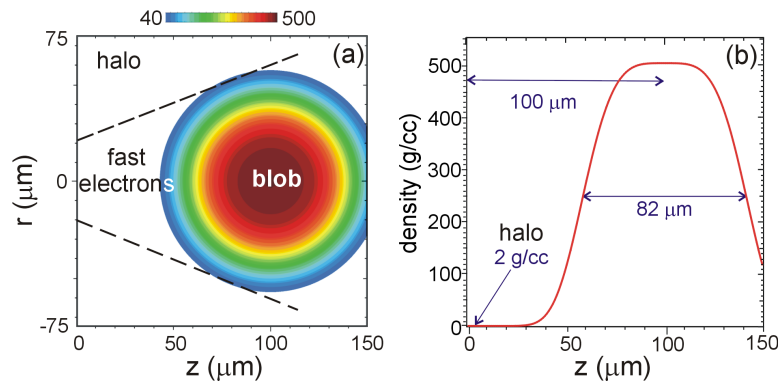}
\end{center}
\caption{\label{fig:2} Central cut through
imploded target configuration: (a) density isocontours
in g/cm$^3$ and (b) density profile at $r$ = 0.}
\end{figure}

\section{Imploded target configuration and beam
parameters.}

The target configuration used in this paper
is shown in figure~\ref{fig:2}. It is a simplified version
of that obtained from radiation-hydrodynamics
calculations \cite{honrubia9, atzeni2}. It consists
of 0.18 mg of DT fuel compressed into a supergaussian
spherical blob with a peak density of 500 g/cm$^3$
and 82 $\mu$m diameter (FWHM) placed on a density
pedestal of 2 g/cm$^3$. The $\rho$R of the blob is
2 g/cm$^2$ and the $\rho$L through the central cut
of the simulation box is 4 g/cm$^2$. For simplicity,
a uniform initial DT temperature of 300 eV has been
assumed, which sets the initial resistivity to a level
of $3\times10^{-8}$ $\Omega$m. We assume also that
the DT is initially at rest, which is a reasonable
choice because the fuel is almost stagnated at the time
of peak $\rho$R.

A beam of fast electrons is injected from the left
at $z=0$. We imagine that it emerges from the tip of
a cone at this position with a supergaussian profile
in radius and Gaussian in time. The supergaussian
profile in radius reduces the beam energy spread
and enhances azimuthal magnetic field growth because
from equation (3)
$\partial B_{\theta}/ \partial t \approx \partial E_z / \partial r
\approx \partial (\eta j_{b,z}) / \partial r$.
However, the lack of experimental evidences of
laser-generated electron beams with such profiles
makes our calculations slightly optimistic in a similar
way to other fast ignition studies that assume
cylindrical beams \cite{solodov2007,atzeni3}. More
realistic calculations with a Gaussian profile have
been presented in Ref. \cite{honrubia5}.

The reference case considered in this paper is defined
by a distance between the cone tip and the dense blob
$d$ = 100 $\mu$m, beam power of  2 PW, beam radius of
20 $\mu$m (HWHM), pulse duration of 18 ps at full width
at half maximum (FWHM), electron mean kinetic energy
within the HWHM of the radial distribution
$\langle E \rangle$ = 1.6 MeV ($f_T$ = 0.67) and initial
beam divergence half-angle $\langle \theta \rangle$
= 35$^{\circ}$ within the HWHM ($f_\theta$ = 1). The
total beam energy is 36 kJ. Assuming a laser-to-fast
electron conversion efficiency of 40\%, this beam
could be generated by a laser pulse with the same
supergaussian distribution and spot radius that the
electron beam, a mean intensity within the FWHM of
2.7$\times$10$^{20}$ W/cm$^{2 }$, $\lambda=$ 0.53 $\mu$m
and a total energy of 90 kJ. The second harmonic of
the neodymium laser is used here to reduce the mean
energy of fast electrons and maximize their coupling
with the dense core. The number density of the electron
beam at the injection surface is around
$2\times 10^{22}$ cm$^{-3}$, below the relativistic
critical density for $I \lambda^2=7.6\times 10^{19}$
W$\mu$m$^2$/cm$^2$, and a factor of $\approx$ 5 higher
than the standard critical density for $\lambda$= 0.53
$\mu$m.

Concerning the numerical parameters, we have chosen a
cell width of 1 $\mu$m in each coordinate, a time step
of 3 fs, and a total number of injected particles of
$4\times10^{6}$. Free boundaries have been used in
all simulations.

\begin{figure}
!\hspace{3.5cm}
\begin{center}
\includegraphics[width=.65\textwidth]{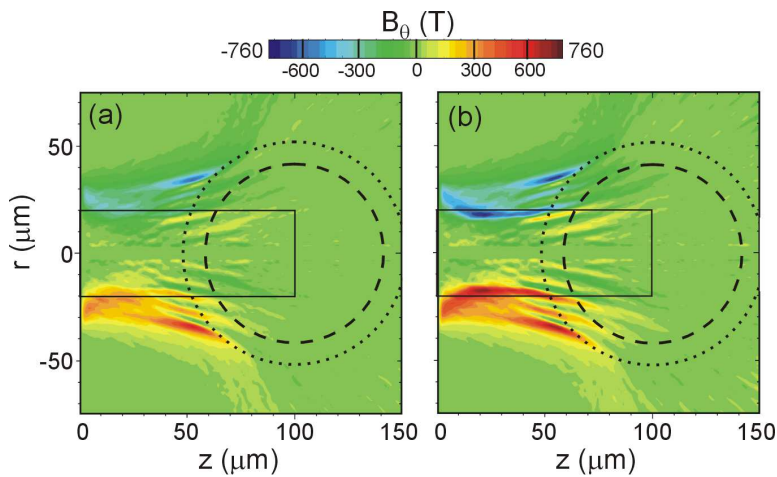}
\end{center}
\caption{\label{fig:3} Self-generated azimuthal
magnetic field $B_{\theta}$ (a) at the time of
peak power and (b) at the end of the pulse.
The fast electron pulse has a Gaussian
profile in time with a total energy of 36 kJ and
a duration of 18 ps (FWHM). Beam parameters are
$d$ = 100 $\mu$m, $\langle E \rangle$ = 1.6 MeV
and $\langle \theta \rangle$ = 35$^{\circ}$.
Dashed and dotted circles show the initial
position of the 250 g/cm$^3$ and 100 g/cm$^3$
isocontours, respectively. The solid line depicts
the position of a perfectly collimated beam of
20 $\mu$m radius.}
\end{figure}

\section{Fast electron energy deposition}

We present in this Section our results on the interaction
and energy deposition of the fast electron beam with the
target for the reference case defined above.

In the low density halo, the beam deposits its energy
mainly as Ohmic heating by return currents. Plasma
electrons are heated up to temperatures $T_e \approx$
10 keV much higher than the ion temperature $T_i$
in a few picoseconds. This temperature is almost
constant while the electron beam is on because the
energy transferred to plasma electrons is balanced
approximately by thermal electron conduction in the
direction perpendicular to the electron beam. The DT
resistivity at 10 keV ($\approx$ 4 $\times $ 10$^{-10}$
$\Omega$m) is high enough for a GA fast electron beam
to generate an azimuthal magnetic field $B_{\theta}$
of hundreds of Tesla in 10 - 15 ps
($B_{\theta} \approx \tau \partial E_z
/ \partial r \approx  \tau \eta j_{b,z} / r_b
= \tau \eta I_b /(\pi r_{b}^3)$,
where $I_b$ stands for total current, $r_{b}$ for beam radius
and $\tau$ for pulse duration). This magnetic field grows up
slowly due to the low plasma resistivity and collimates the
electron beam in the second half of the pulse. Azimuthal
magnetic fields of a few hundreds of Tesla are sufficient
to collimate most of the beam electrons. For instance,
the Larmor radius of 1 MeV electron in a $B_{\theta}$-field
of 250 T is around 19 $\mu$m, similar to the 20 $\mu$m beam
radius assumed here. If thermal conduction were not taken
into account, temperatures in the halo would be much higher,
the $B_{\theta}$-field would saturate at levels lower than
the hundreds of Tesla reported here and there would be no
significant beam collimation in the halo, as it was shown
in previous studies \cite{honrubia4}.

At intermediate densities 10 $<$ $\rho$ $<$
100 g/cm$^3$, electron
temperatures are lower, resistivities higher and
electron-ion energy exchange is fast enough to
equilibrate electron and ion temperatures $T_i \approx T_e$.
Higher resistivities lead to enhanced field generation
and collimation of the relativistic electron
beam after a few ps, as depicted in figure~\ref{fig:3}.
It is worthwhile noticing that the $B_{\theta}$-field
peaks in this zone and how its oscillations are
attenuated towards the beam axis due to the increasing
plasma density. The enhancement of the beam collimation
with time can be observed by comparing the location
of the peak $B_{\theta}$-field, which is near
the beam edge, in Figs.~\ref{fig:3}(a) and (b).

\begin{figure}
!\hspace{4cm}
\begin{center}
\includegraphics[width=.58\textwidth]{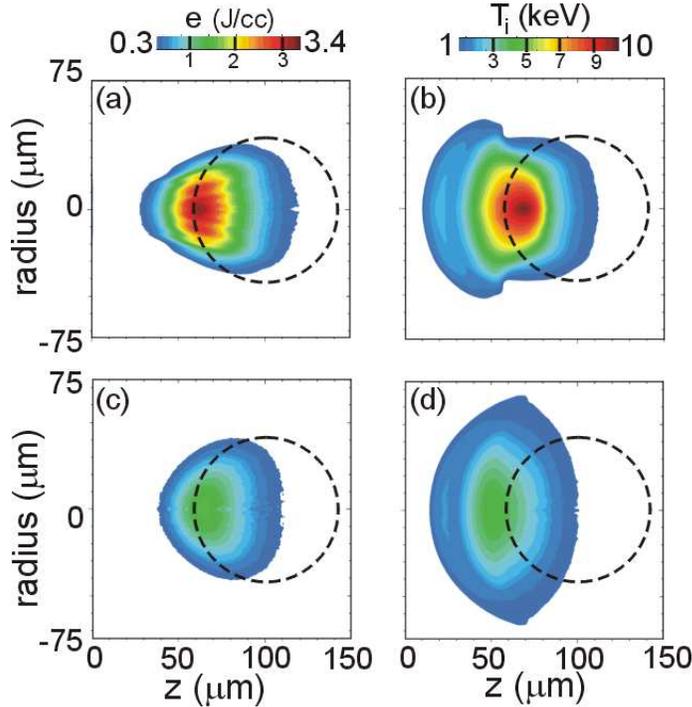}
\end{center}
\caption{\label{fig:4} Energy density in J/cm$^{3}$
 deposited by the fast electron beam for the case of
 figure \ref{fig:3}. Upper row [(a) and (b)]:
 Coulomb energy deposition and ion temperature at the
 end of the pulse in the full simulation with
 self-generated fields on. Lower row [(c) and (d)]
 Coulomb energy deposition and ion temperature at the
 end of the pulse in the simulation with self-generated
 fields artificially suppressed. Dashed circles show
 the initial position of the 250 g/cm$^3$
 density isocontour.}
\end{figure}

\begin{figure}
!\hspace{5cm}
\begin{center}
\includegraphics[width=.55\textwidth]{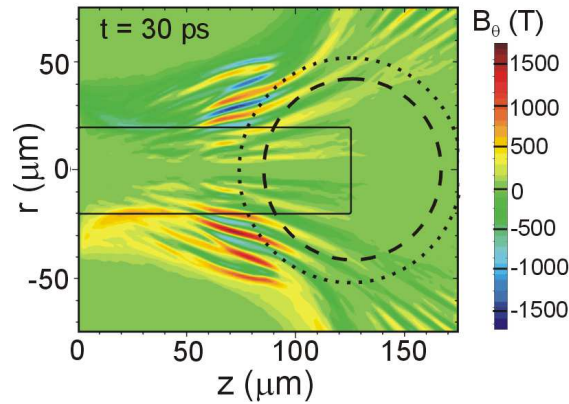}
\end{center}
\caption{ \label{fig:5} 
Self-generated azimuthal magnetic field B$_{\theta}$ for the
target pictured in figure \ref{fig:2} with $d$ = 125 $\mu$m.
The beam parameters are $\langle E \rangle$ = 1.6 MeV,
$\langle \theta \rangle$ = 35$^{\circ}$ and a total energy
of 50 kJ. Dashed and dotted circles show the initial position
of the 250 g/cm$^3$ and 100 g/cm$^3$ isocountours, respectively.
The solid line depicts the position of a perfectly collimated
beam  of 20 $\mu$m radius.}
\end{figure}

At densities higher than 100 g/cm$^3$,
self-generated fields are damped by
collisions of beam electrons with the background
plasma and DT heating is almost exclusively due to
Coulomb deposition. As has been pointed out
recently \cite{honrubia4}, anomalous electron stopping
due to resistivity effects plays no significant
role at such high densities. The overall energy
balance of the reference case shows that 87\%
of the beam energy is deposited in the target
via Coulomb collisions, 12\% is carried by electrons
that pass through the target and escape, 1\% is
deposited as Ohmic heating  by return currents
and around 54\% is deposited at densities
higher than half of the maximum
($\rho$ $>$ 250 g/cm$^3$). Even though 
only 1\% of the beam energy is deposited as Ohmic
heating, self-generated fields turn out to
contribute to the Coulomb energy deposition in the
dense core indirectly via resistive collimation
of the fast electron beam, which concentrates
the energy deposition in a spot slightly bigger
than the initial beam diameter, as shown in
figure~\ref{fig:4}(a). The effective propagation
full-angle of the beam electrons near the dense
core taken from Figs.~\ref{fig:3} and \ref{fig:4}
is, approximately, 28$^{\circ}$, which gives a maximum
spot diameter of 49 $\mu$m (FWHM) at a depth of
$z$ = 72 $\mu$m. The energy density deposited via
Coulomb collisions with and without self-generated
fields are compared in Figs.~\ref{fig:4}(a) and (c).
It is worth mentioning
the enhancement of the energy density by a factor of
about 2  when self-generated fields are taken into
account. The oscillations shown in figure~\ref{fig:4}(a)
are due to the weak filamentation of the $B_\theta$-field
in the density ramp. These oscillations do not affect
DT ignition significantly because of thermal smoothing.
However, for higher distances $d$ from cone to dense core,
the beam becomes more filamented (see figure~\ref{fig:5})
and the energy deposition more fragmented than
that shown in figure~\ref{fig:4}(a), imprinting the
plasma density and temperature profiles.

\section{Coupling efficiency}

\begin{figure}
\begin{center}
\includegraphics[width=0.95\textwidth]{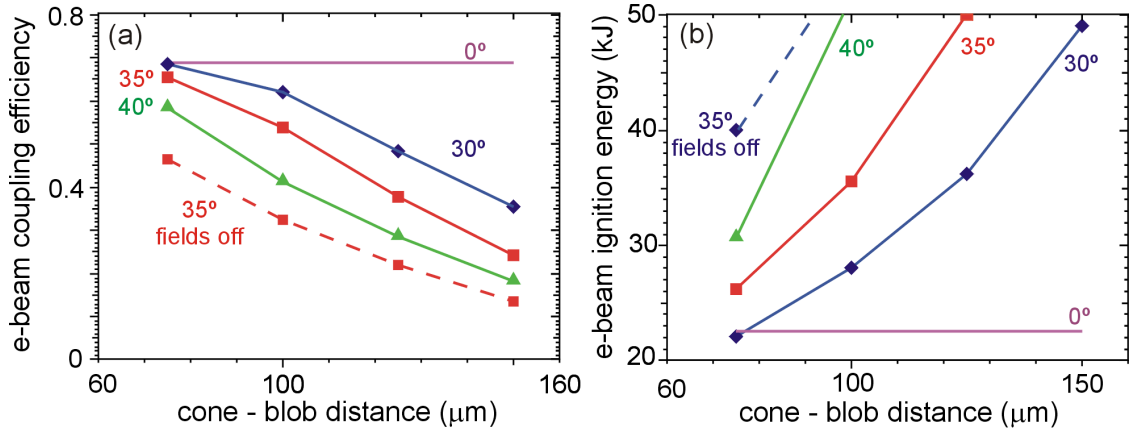}
\end{center}
\caption{\label{fig:6} (a) Coupling efficiencies and (b)
minimum ignition energies of electron beams with
$\langle E \rangle$ = 1.6 MeV as function of the
cone-blob distance. Curves are labelled with the
initial beam divergence half-angle $\langle \theta \rangle$. 
The dashed lines and the lines labelled with 0$^{\circ}$
correspond to simulations with self-generated fields
artificially suppressed.}
\end{figure}

The electron beam coupling efficiencies (defined as
the fraction of the beam energy deposited at densities
higher than 250 g/cm$^3$) predicted by our integrated
model are depicted in figure~\ref{fig:6}(a). The corresponding
electron-beam ignition energies $E_{ig}$ are shown in
figure~\ref{fig:6}(b). In all simulations we keep constant
the fast electron beam intensity and increase the beam
energy firstly by increasing the pulse duration up to
a maximum of 20 ps and then by increasing the beam
radius. For instance, the ignition energy of 57 kJ is
obtained for a pulse duration of 20 ps and a beam
radius of 23.9 $\mu$m  while these parameters are
18 ps and 20 $\mu$m for the reference case of 36 kJ.

It is remarkable that, for the beam parameters chosen
here, fast electrons propagate up to the top of the
density ramp and deposit there a significant fraction
of their energy without beam disruption or breaking-up
due to the self-generated magnetic fields. Beams with
lower kinetic energies are more prone to filamentation,
while beams with higher kinetic energies have lower
coupling efficiencies \cite{honrubia4}. The importance
of beam collimation can be seen in
figure~\ref{fig:6}(a) by comparing the coupling
efficiencies for different $d$ and $\langle \theta \rangle$
with that obtained for a perfectly collimated beam
($\langle \theta \rangle$ = 0$^{\circ}$ and fields
turned off). As shown in the figure, only
for $d$ = 75 $\mu$m and $\langle \theta \rangle$ =
30$^{\circ}$ the beam is nearly collimated, decreasing
beam collimation with divergence angles, in agreement
with the results of Bell and Kingham \cite{bell2}.

The dependence of the coupling efficiency on the mean kinetic
energy of beam electrons is shown in figure~\ref{fig:7}.
If self-generated fields are artificially suppressed, the optimal
kinetic energy changes with the initial divergence half-angle
from around 2 MeV for perpendicular incidence to about 1.2 MeV
for $\langle \theta \rangle$ = 35$^{\circ}$. This is due to
the different target areal densities seen by fast electrons
with different divergence angles. In full simulations with
fields on and $\langle \theta \rangle$ = 35$^{\circ}$,
the coupling efficiencies are much higher due to beam
collimation and have a maximum around 1.2 MeV, similar
to that found in simulations with fields off. It is
shown also in this curve the weak dependence of the
coupling efficiency on mean kinetic energy for
$\langle E \rangle \le$ 1.6 MeV. For higher kinetic
energies, beam current density, self-generated fields
and beam collimation are lower and the coupling
efficiency tends to that found with fields off.

Assuming a laser-to-fast electron conversion
efficiency of 40\% and a minimum overall coupling efficiency of the
laser beam to the dense core of 25\% \cite{atzeni2}, suitable
fast ignition schemes have to have minimum electron beam coupling
efficiencies around 0.6. Figure~\ref{fig:6}(a) shows that
this coupling efficiency is found for $\langle \theta \rangle
\le$ 35$^{\circ}$ and distances $d$ $\le$100 $\mu$m, approximately,
for the target analysed here. Figure~\ref{fig:7} adds the additional
restriction that $\langle E \rangle \le$ 1.6 MeV to achieve overall
coupling efficiencies around 25\%.

\section{Fuel ignition}

\begin{figure}
!\hspace{5cm}
\begin{center}
\includegraphics[width=.5\textwidth]{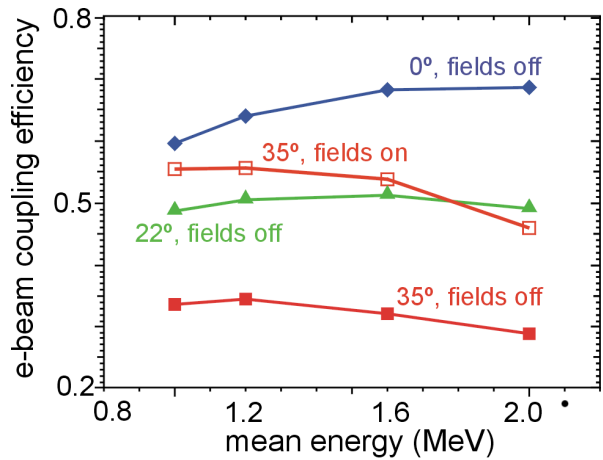}
\end{center}
\caption{\label{fig:7} Electron beam coupling efficiency
as function of the mean kinetic energy of the beam electrons.
Curves with solid markers correspond to simulations with Coulomb
energy deposition only. The curve with open squares corresponds
to full simulations including self-generated fields. Each curve is labelled
with the initial beam divergence half-angle. In all cases the distance
between the cone and the blob is $d$ = 100 $\mu$m.}
\end{figure}

\begin{figure}
!\hspace{3.5cm}
\begin{center}
\includegraphics[width=.7\textwidth]{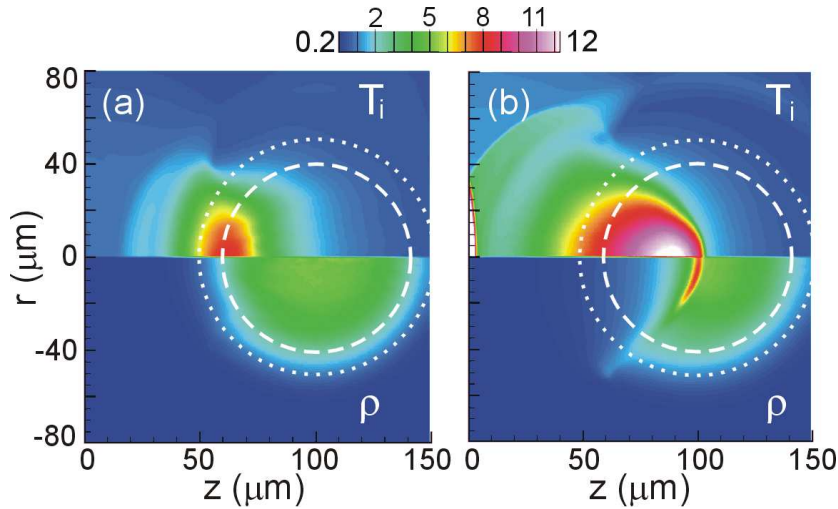}
\end{center}
\caption{\label{fig:8} 
Ion temperature and plasma density isocontours for the case
(a) and (b) of Figs. \ref{fig:4} with self-generated fields on
(a) when 90\% of the beam energy has been injected
in the simulation box, and (b) 30 ps after the time of
peak power of the fast electron beam when ignition is propagating
through the DT fuel. Dashed and dotted circles show the initial
position of the 250 g/cm$^3$ and 100 g/cm$^3$ density
isocontours, respectively. Ion temperatures are given
in keV and densities in units of 100 g/cm$^3$.}
\end{figure}

We have performed a series of integrated simulations to obtain
the ignition energies $E_{ig}$ as a function of cone to blob
distance $d$, initial divergence half-angle $\langle \theta \rangle$
and mean kinetic energy of fast electrons $\langle E \rangle$.
Typical density and ion temperature profiles near the
end of the pulse and in the ignition propagation phase are
depicted in figure~\ref{fig:8}, where one can see that DT
expansion is not important during the beam energy deposition
and becomes relevant after fusion reactions have been triggered.
The ignition energies shown in this Section are obtained
as the minimum beam energy for which the thermonuclear fusion
power has an exponential or higher growth in time. Ignition
energies as a function of $d$ and $\langle \theta \rangle$
for a beam with $\langle E \rangle$ = 1.6 MeV are plotted in
figure~\ref{fig:6}(b). This figure shows the very important
dependence of $E_{ig}$ on both parameters and the crucial
role played by self-generated fields. The ignition energies
of figure~\ref{fig:6}(b) are 20-30\% lower than those
reported in previous studies \cite{honrubia5} due to
the sharper radial profile of the fast electron beam
assumed here, i.e. supergaussian versus Gaussian, which
concentrates the energy deposition and enhances magnetic
field generation at the beam edge. Comparison of the
ignition energies with that obtained for a perfectly
collimated beam ($\langle \theta \rangle$ = 0$^{\circ}$ and
fields turned off) shows that for $d$ = 75 $\mu$m and
$\langle \theta \rangle$ = 30$^{\circ}$, the beam is almost
perfectly collimated. For higher $d$ and $\langle \theta \rangle$,
beam collimation is still important, but the beam diverges with
an angle lower than the initial divergence assumed when propagates
toward the blob.

\begin{figure}
!\hspace{5cm}
\begin{center}
\includegraphics[width=.50\textwidth]{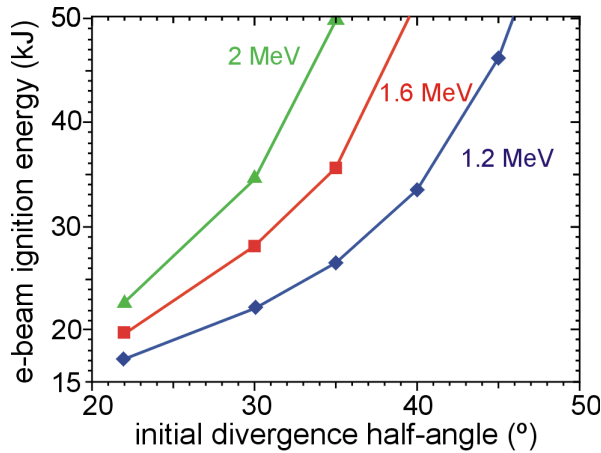}
\end{center}
\caption{\label{fig:9} Electron beam ignition energies 
of the target depicted in figure \ref{fig:2}
as a function of the initial divergence
half-angle $\langle \theta \rangle$ with the
mean kinetic energy $\langle E \rangle$ as a
parameter. The distance between cone and blob
is $d$ = 100 $\mu$m in all curves.}

\end{figure}

Sensitivity of the ignition energies on
$\langle \theta \rangle$ and $\langle E \rangle$ is
shown in figure~\ref{fig:9} for a target with
$d$ = 100 $\mu$m. The relatively low sensitivity found
for low initial divergences is a signal of beam collimation.
For instance, the ignition energy $E_{ig}$ = 19.5 kJ obtained
for a beam with $\langle E \rangle$ = 1.6 MeV and
$\langle \theta \rangle$ = 22${^\circ}$ is lower than that
obtained for a perfectly collimated beam (22.5 kJ in
figure~\ref{fig:6}(b)) due to the beam compression by the
$B_{\theta}$-field. It is also worth pointing out that
the lowest ignition energies are obtained for
$\langle E \rangle$ = 1.2 MeV, in agreement with the
coupling efficiencies shown in figure~\ref{fig:7}.
However, the dependence of $E_{ig}$ on $\langle E \rangle$
is more pronounced than that found for the coupling
efficiency due to the variation of the penetration
depth with the mean kinetic energy of beam electrons.

Our integrated simulations predict that the imploded
target configuration pictured in figure~\ref{fig:2}
heated by an electron beam with $\langle E \rangle$
= 1.6 MeV and $\langle \theta \rangle$ = 30$^{\circ}$
will ignite with beam energies $E_{ig}$ between 22
and 50 kJ, depending of the distance $d$ between the
cone and the blob. For the laser-to-fast-electron
conversion efficiency of 0.4 assumed, these energies
correspond to laser pulses from 55 to 125 kJ, respectively.
Ignition of targets with $d$ = 100 $\mu$m by laser beam
energies lower than 100 kJ requires to generate electron
beams with initial divergences and mean energies within
the zone defined by $E_{ig} \le$ 40 kJ in figure~\ref{fig:9},
which sets a compromise between the beam parameters
$\langle E \rangle$ and $\langle \theta \rangle$ for
the imploded target configuration analysed here.

\section{Conclusions}

One of the main conclusions of our study is that giga-ampere, multi-PW
currents can be transported through the steep gradients of the
plasma corona toward the high-density fuel core and can ignite
it with beam energies of 30 - 40 kJ for the electron beam parameters
and the target configuration assumed. The present simulations 
show that in the dense core, energy deposition takes place
almost exclusively by classical Coulomb collisions. Self-generated
fields play a major role for core heating improving the coupling
efficiency substantially, but in an indirect way by means of beam
collimation.

We have performed a parametric study to obtain minimum ignition
energies as a function of the cone - blob distance, beam divergence
and electron kinetic energy for the imploded fuel configuration
considered here. We found that the ignition energy depends strongly
on the cone - blob distance and the beam divergence, both crucial
parameters for fast ignition. The optimal kinetic energy for the initial
divergence half-angle of 35$^{\circ}$ is about 1.2 MeV, significantly
lower than the 2 MeV found for perfectly collimated beams. Our
calculations show that targets with cone - blob distances lower
than 125 $\mu$m can be ignited by electron beams with energies
around 40 kJ if initial divergence half-angles are about 30$^{\circ}$
- 35$^{\circ}$ and electron kinetic energies are lower than
1.6 MeV. Assuming a laser-to-fast electron conversion efficiency
of 40\%, these electron beams could be generated by the short-pulse
laser beams around 100 kJ envisioned for future facilities
\cite{dunne}. These conclusions rely on the integrated hybrid PIC
model used here, which is valid for electron transport in dense media
and therefore depends on the distribution function assumed for the
injected electrons. Further theoretical and experimental investigations
for a full characterization of fast electron beams generated in cones
by laser pulses of 10 - 20 picoseconds are necessary to estimate more
precisely the energy requirements of electron-driven fast ignition.

\section*{Acknowledgments}
This work was supported by the research grants ENE2006-06339
and CAC-2007-13 of the Spanish Ministry of Education and by the Association
EURATOM - IPP Garching in the framework of IFE Keep-in-Touch
Activities and the Fusion Mobility Programme.

\newpage


\section*{References}
\begin{thebibliography}{10}

\bibitem{tabak1} Tabak M, Hammer J, Glinsky M, Kruer W L, Wilks S C,
	Woodworth J, Campbell E M, Perry M D and Mason R J 1994 {\it Phys. Plasmas} {\bf 1} 1626 

\bibitem{tabak2} Tabak M, Clark D S, Hatchett S P, Key M H, Lasinski B F, Snavely R A,
	Wilks S C, Town R P J, Stephens R, Campbell E M, Kodama R, Mima K, Tanaka K A
	Atzeni S and Freeman R 2005 {\it Phys. Plasmas} {\bf 12} 052708

\bibitem{honrubia4} Honrubia J J and Meyer-ter-Vehn J 2006 {\it Nucl. Fusion} {\bf 46} L25

\bibitem{honrubia5} Honrubia J J and Meyer-ter-Vehn J 2008 {\it Journal of Physics: Conference Series}
	{\bf 112} 022055

\bibitem{Green}  Green J S, Ovchinnikov V M, Evans R G, Akli K U, Azechi H, Beg F N, Bellei C, Freeman R R,
    	Habara H, Heathcote R, Key M H, King J A, Lancaster K L, Lopes N C, Ma T, MacKinnon A J,
     	Markey K, McPhee A, Najmudin Z, Nilson P, Onofrei R, Stephens R, Takeda K, Tanaka K A,
	Theobald W, Tanimoto T, Waugh J, Van Woerkom L, Woolsey N C, Zepf M, Davies J R and Norreys P A
	2008 {\it Phys. Rev. Lett.} {\bf 100} 015003

\bibitem{Lancaster} Lancaster K L, Green J S, Hey D S, Akli K U, Davies J R, Clarke R J,Freeman R R,
	Habara H, Key M H, Kodama R, Krushelnick K, Murphy C D, Nakatsutsumi M, Simpson P, Stephens R,
	Stoeckl C, Yabuuchi T, Zepf M and Norreys P A 2007 {\it Phys. Rev. Lett.} {\bf 98} 125002

\bibitem{sentoku2008} Chrisman B, Sentoku Y and Kemp A J 2008 {\it Phys. Plasmas} {\bf 15} 056309

\bibitem{dunne} Dunne M 2006 {\it Nature Phys.} {\bf 2} 2

\bibitem{solodov2007} Solodov A A, Betti R, Delettrez J A and Zhou C D 2007 {\it Phys. Plasmas} {\bf 14}, 062701

\bibitem{bell} Bell A R, Davies J R, Guerin S and Ruhl H 1997 {\it Plasma Phys. Control. Fusion} {\bf 39} 653

\bibitem{davies} Davies J R 2002 {\it Phys. Rev. E} {\bf 65} 026407

\bibitem{gremillet} Gremillet L , Bonnaud G and Amiranoff F 2002 {\it Phys. Plasmas} {\bf 9} 941

\bibitem{kemp2} Kemp A J, Sentoku Y, Sotnikov V and Wilks S C 2006 {\it Phys. Rev. Lett.} {\bf 97} 235001

\bibitem{kodama1} Kodama R, Norreys P A, Mima K, Dangor A E, Evans R G, Fujita H, Kitagawa Y, Krushelnick K,
	Miyakoshi T, Miyanaga N, Norimatsu T, Rose S J, Shozaki T, Shigemori K, Sunahara A, Tampo M,
	Tanaka K A, Toyama Y, Yamanaka T and Zepf M 2001 {\it Nature} {\bf 412} 798

\bibitem{kodama2} Kodama R, Shiraga H, Shigemori K, Toyama Y, Fujioka S, Azechi H, Fujita H, Habara H, Hall T,
	Izawa Y, Jitsuno T, Kitagawa Y, Krushelnick K M, Lancaster K L, Mima K, Nagai K, Nakai M, Nishimura H,
	Norimatsu T, Norreys P A, Sakabe S, Tanaka K A, Youseff A, Zepf M and Yamanaka T 2002 {\it Nature} {\bf 418} 933

\bibitem{campbell} Campbell R B, Kodama R, Mehlhorn T A, Tanaka K A and Welch D R 2005 {\it Phys. Rev. Lett.} {\bf 94} 055001

\bibitem{mason} Mason R J 2006 {\it Phys. Rev. Lett.} {\bf 96} 035001

\bibitem{honrubia2} Honrubia J J, Antonicci A and Moreno D 2004 {\it Las. Part. Beams} {\bf 22} 129

\bibitem{honrubia1} Honrubia J J, Kaluza M, Schreiber J, Tsakiris G D and Meyer-ter-Vehn J 2005 {\it Phys. Plasmas} {\bf 12} 052708

\bibitem{honrubia3} Honrubia J J, Alfonsín C, Alonso L, Perez B, and Cerrada J A 2006 {\it Las. Part. Beams} {\bf 24} 217

\bibitem{Sentoku2004} Sentoku Y, Mima K, Ruhl H, Toyama Y, Kodama R and Cowan T E 2004 {\it Phys. Plasmas} {\bf 11} 3083

\bibitem{silva} Silva L O, Fonseca R A, Tonge J W, Mori W B and Dawson J M 2002 {\it Phys. Plasmas} {\bf 9} 2458

\bibitem{bret} Bret A, Firpo M C and Deutsch C 2005 {\it Phys. Rev. Lett.} {\bf 94} 115002

\bibitem{pukhov} Pukhov A, Sheng Zh-M and Meyer-ter-Vehn J 1999 {\it Phys. Plasmas} {\bf 6} 2847

\bibitem{nakamura2008}  Nakamura T, Sentoku Y, Matsuoka T, Kondo K, Nakatsutsumi M,
	Norimatsu T, Shiraga H, Tanaka K A, Yabuuchi T and Kodama R 2008 {\it Phys. Rev. Lett.}	{\bf 100} 165001

\bibitem{nakamura2004}  Nakamura T, Kato S, Nagatomo H and Mima K 2004 {\it Phys. Rev. Lett.} {\bf 93} 265002

\bibitem{nakamura2007}	Nakamura T, Mima K, Sakagami H and Johzaki T 2007 {\it Phys. Plasmas} {\bf 14} 053112

\bibitem{Baton} Baton S D, Koenig M, Fuchs J, Benuzzi-Mounaix A, Guillou P, Loupias B, Vinci T, Gremillet L,
	Rousseaux C, Drouin M, Lefebvre E, Dorchies F, Fourment C, Santos J J, Batani D, Morace A, Redaelli R,
	Nakatsutsumi M, Kodama R, Nishida A, Ozaki N, Norimatsu T, Aglitskiy Y, Atzeni S and Schiavi A 2008
	{\it Phys. Plasmas} {\bf 15} 042706

\bibitem{Salvat} Sempau J, Fernández-Varea J M, Acosta E and Salvat F 2003 {\it Nucl. Instr. Methods in Phys. Res. B} {\bf 207} 107

\bibitem{Li} Li C K and Petrasso R D 2006 {\it Phys. Plasmas} {\bf 13} 056314

\bibitem{solodov}  A. A. Solodov and R. Betti, {\it Phys. Plasmas} {\bf 15}, 042707 (2008)

\bibitem{deutsch} Starikov K V and Deutsch C 2005 {\it  Phys. Rev. E} {\bf 71} 026407

\bibitem{deutsch1} Deutsch C, Furukawa H, Mima K, Murakami M and Nishihara K 1996 {\it Phys. Rev. Lett.} {\bf 77} 2483

\bibitem{atzeni3} Atzeni S, Schiavi A and Bellei C 2007 {\it Phys. Plasmas} {\bf 14} 052702

\bibitem{NRL} Huba J D 1994 {\it NRL Plasma Formulary}, Naval Research Laboratory, Washington DC

\bibitem{kemp} Kemp A J and Meyer-ter-Vehn J 1998 {\it Nucl. Instr. Methods in Phys. Res. A} {\bf 415} 674

\bibitem{honrubia9} Honrubia J J, unpublished simulation of cone-targets with the code SARA-2D. See also \cite{atzeni2}.

\bibitem{atzeni2} Atzeni S, Schiavi A, Honrubia J J, Ribeyre X, Schurtz G, Nicolaï Ph, Olazabal-Loumé M, Bellei C,
	Evans R G and Davies J R 2008 {\it Phys. Plasmas} {\bf 15} 056311

\bibitem{bell2} Bell A R and Kingham R J 2003 {\it Phys. Rev. Lett.} {\bf 91} 035003


\end{thebibliography}
\end{document}